\documentclass[review]{elsarticle}

\usepackage{lineno,hyperref}
\usepackage{color}
\usepackage{amssymb}
\usepackage{graphics}
\usepackage{graphicx}
\usepackage{pbox}
\usepackage{mhchem}

\usepackage{xcolor} 
\usepackage[ampersand]{easylist}
\usepackage{lineno}
\usepackage{booktabs}
\usepackage{multirow}
\usepackage{subcaption}
\usepackage{amsmath}
\usepackage{paralist}
\usepackage{soul}
\usepackage{siunitx}
\usepackage{amsmath}
\usepackage{url}
\usepackage{makecell}

\usepackage{hyperref}

\hypersetup{  
    colorlinks=true,
    linkcolor=blue,
    filecolor=magenta,      
    urlcolor=cyan,
    pdftitle={Sharelatex Example},
    bookmarks=true,
    pdfpagemode=FullScreen,
    citecolor=blue
    }

\journal{Nuclear Instruments and Methods in Physics Research A }

\begin{document}

\begin{frontmatter}

\title{Hydrogenous content identification in heterogeneous cargoes via multiple monoenergetic neutron radiography}

\author[MITaddress]{Jill Rahon
}
\author[MITaddress]{Areg Danagoulian\corref{corauthor}}
\cortext[corauthor]{Corresponding author}
\ead{aregjan@mit.edu}

\address[MITaddress]{Massachusetts Institute of Technology, Cambridge, MA 02139, USA}

\begin{abstract}

The determination of hydrogenous content in commercial cargoes is an important challenge in cargo security.
Prior work has shown the feasibility of hydrogenous cargo classification in radiographic applications.  This result was achieved by using the significant material to material differences in the energy dependence of their neutron scattering cross-sections. 
The work presented here details the application of this technique to multiple-monoenergetic neutron transmission measurements of several heterogeneous cargo mock-ups with the goal of quantifying hydrogenous content. It demonstrates the ability to determine the hydrogenous content of a cargo container by performing
 analysis of pulse-height data of transmitted neutrons. The set-up used for these feasibility studies was able to detect and quantify areal densities of up to 30 g/cm$^2$ of high density polyethylene (HDPE), even when mixed with metallic materials. Accurate determination of the hydrogenous content of the cargo has two important uses:  it can allow for better imaging and discrimination of conventional contraband during cargo screening; it may also allow the unfolding of the individual elemental contributions from an effective atomic number, $Z_{\mathrm{eff}}$, as determined from multiple monoenergetic gamma radiography (MMGR).  This combination of capabilities could make multiple monoenergetic neutron radiography applications a significant contribution to cargo security. 
\end{abstract}

\begin{keyword}
Active interrogation \sep Nuclear security \sep Radiography
\end{keyword}
\end{frontmatter}


\section{Introduction}

Every day approximately 50 000 maritime ISO containers enter the United States~\cite{kouzes}.  This flow of cargoes currently undergoes a variety of checks for the presence of nuclear and conventional contraband.  
Significant efforts have taken place over the last two decades on the development of passive, active, and radiographic screening systems.  The goal of these systems is to detect and identify a variety of threats:  radioactive materials;  various Special Nuclear Materials (SNM);  fully assembled nuclear devices, whose danger is significant in the contect of nuclear terrorism; and conventional contraband, which is smuggled for a variety of economic reasons. Passive techniques primarily focus on the detection of the natural emission of particles from the radioactive materials.  Active techniques involve irradiation of the cargo with various particles with the purpose of triggering element-specific processes, whose detection by proxy will allow the inference of the cargoe's elemental composition.  Such techniques involve detection of photofission neutrons and photons, detection of secondary bremsstrahlugn photons, nuclear resonance fluorescence (NRF),  etc~\cite{pnpf-short,PNPF_patent,ref:bertozzi}.   Radiographic techniques, on the other hand, use measurements of transmitted particle beams to infer the areal density and in some cases, the effective atomic number $Z_{\mathrm{eff}}$ of the cargo~\cite{MMGR,rose2016}.  Imaging of the cargo based on $Z_{\mathrm{eff}}$, in addition to its utility in locating actinides, can also be used to differentiate between medium-Z and low-Z cargoes, thus providing additional information to the operators in their effort to detect various contraband.  The transmission of neutron beams have been used in various configurations in combination with photon transmission to estimate the areal density and composition of the cargoes beyond what can be done with photons alone~\cite{ref:sowerby2007recent,ref:blackburn2007fast,nattress2016,van,sowerby}.  In particular, work by Cutmore {\it et al.} shows that neutron counting along with photon radiography can allow the imaging and improved identification of a variety of hydrogenous materials\cite{cutmore}.  These methods, however, use only neutron counting, and do not exploit the spectral information of the transmitted neutron flux.  Additional work by others has showing that precise reconstructions of neutron energy can allow for resonance radiography applications which allow for the determination of carbon, oxygen, and nitrogent contents of the cargo~\cite{chen2002fast,perticone2019fast,mor2015reconstruction,overley2006explosives}. Our work extends this by showing that the analysis of the neutron deposited energy alone, which results in a much simpler measurements, can provide information about the hydrogenous content.  It can then further improve the cargo content characterization and thus provide additional information not available in simple neutron counting based methods.  It should be added that the work presented here details a new technique which, rather than as a standalone system, can be combined with other radiographic techniques to achieve the above mentioned improvements.  While the system described in this work uses a Radio Frequency Quadrupole (RFQ) as a platform and the $^{11}B(d,n\gamma)^{12}C$ reaction as a source of neutrons and photons, essentially any source of polychromatic neutrons could be used, as long as it spans the neutron energy range of $\sim [1,10]$~MeV, where the variation of fast neutron scattering cross sections on hydrogen is the greatest.  Examples of such platforms could be:
\begin{itemize}
    \item A combination of deuteron-deuteron (DD) and deuteron-triton (DT) sources, or a DT source with its 14 MeV neutron energy broadened with (n,2n) reactions.  More specifically, the technique described in Ref.~\cite{cutmore} could be augmented with an additional DD source, thus allowing to perform differential neutron transmission measurements similar to the technique presented in this work.
    \item an isotopic neutron source, such as $^{252}$Cf source
    \item a source of neutrons from an 9 MeV X-ray scanner fitted with a berylium neutron converter, producing neutrons in the [0, 6.5] MeV range via the $^9Be(\gamma,n)^8Be$ reaction.
\end{itemize}    

This work builds upon and expands on the results achieved as part of feasibility studies using multiple monoenergetic neutrons from the $^{11}B(d,n\gamma)^{12}C$ reaction~\cite{ref:rahon2016spectroscopic}.  The previous work showed that it is possible to use spectroscopic analysis of transmitted multiple monoenergetic neutrons to estimate the hydrogenous content of a homogeneous cargo.  Similar work done by other groups since the publication of these studies have shown that the basic concept behind the technique can be scaled to a multi-sensor system~\cite{ref:sweany2016design}.  The original interest in the $^{11}B(d,n\gamma)^{12}C$ reaction, which is excited via a 3 MeV deuteron beam, focused on the 4.4 MeV and 15.1 MeV photons which it produces.  These photons can be used for performing dual energy radiographic measurements which allow the determination of the $Z_{\mathrm{eff}}$ of the cargo via multiple monoenergetic gamma radiography (MMGR)~\cite{MMGR}.  However, such a determination is limited when it comes to SNM detection. In a simple, non-tomographic radiography a given value of $Z_{\mathrm{eff}}$ is not specific enough to determine a cargo type:  various combinations of different elements could produce the same $Z_{\mathrm{eff}}$.  This circumstance could enable the smuggler to combine uranium with hydrogenous material (e.g. plastic, wood, water, etc) to reduce the effective $Z$ of the cargo and thus avoid interception.  

This circumstance creates the need for additional modalities for the detection of SNM or clearing cargoes of their presence.  Multiple monoenergetic neutron radiography is uniquely sensitive to hydrogen content and given the multi-particle nature of the reaction, it requires no additional source.  This work focuses on the study of neutron radiography of heterogeneous cargoes, where various amounts of high density polyethylene (HDPE) were placed behind metallic cargo objects of various thickness and atomic number.  The analysis of transmission data via the simple algorithm described in Ref.~\cite{ref:rahon2016spectroscopic} shows that the technique can detect hydrogenous content, even in small concentrations.  Furthermore, this analysis can estimate the HDPE fraction in a given pixel, which, if combined with MMGR data on $Z_{\mathrm{eff}}$, can result in a combined detection system of a higher specificity.  Though the described spectral analysis technique is based on neutrons generated by a single nuclear reaction, it can be applied in any apparatus using dual or multi energy neutron groups.

\section{Concept}\label{sec:concept}

\begin{figure}
    \centering
    \includegraphics{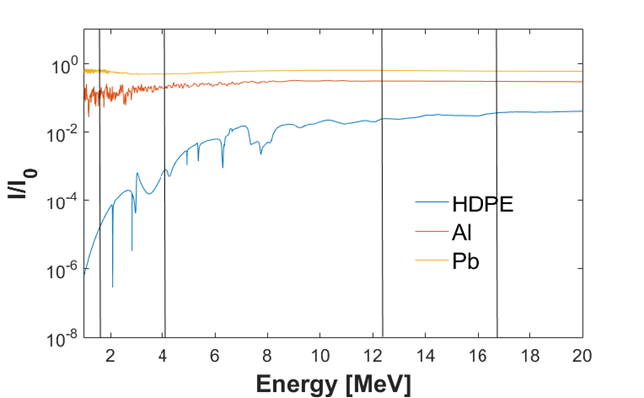}
    \caption{The neutron transmission ratio for various materials.  The presence of hydrogen in HDPE causes a strong variations in transmission amplitude across the MeV energies. Vertical lines indicate discrete neutron energies generated by 3.0 MeV deuteron in the $^{11}B(d,n)^{12}C$ reaction.}
    \label{fig:ii0}
\end{figure}

In standard transmission radiography, total counts of transmitted neutrons are
compared with the known flux of the incident beam to generate a ratio representative
of neutron interaction within the scanned material.  
The cross section, $\sigma$, for neutrons of a defined energy is a constant for a given isotope and type of interaction. This cross section can vary dramatically with neutron energy, as is the case for the total $\sigma_n(E)$ for hydrogen and neutron energies between 1 and 15 MeV. This shift results in a significant change in neutron beam attenuation in hydrogenous materials, such as polyethylene, based on incident particle energy. The final intensity, $I$ of a neutron beam, transmitted through a mixed element target of thickness $x$ and density $\rho$ is calculated by 

\begin{equation}\label{eq:attenuation}
I = I_0 \text{exp}\Big\{-\rho x N_A  \frac{\sum_i c_i \sigma_i(E)}{\sum_i c_i A_i} \Big\},
\end{equation}
where $I_0$ is the incident flux, the sum is over the $i$ elements in the material, $N_A$ is Avogadro's number, $\sigma_i$ is the energy dependent total neutron interaction cross section for element $i$, $c_i$ is the elemental fraction in the molecule of the material, and $A_i$ is the atomic weight, respectively.  Eq.~\ref{eq:attenuation} can be rewritten as

\begin{equation}\label{eq:attn_short}
\frac{I}{I_0} = e^{{-\Sigma_{tot}x}}
\end{equation}
where $\Sigma_{tot}=N_A\rho \frac{\sum_i c_i \sigma_i(E)}{\sum_i c_i A_i}$ is the total macroscopic cross section for all neutron interactions which effectively remove particles from the beam. Integrating the transmitted neutron counts, $I$, over a particular energy region gives the total count of non-attenuated monoenergetic neutrons. Comparing this value for two energy regions yields an indication of the $\sigma_n$ of the transited material. This ratio, $S$, defined as
\begin{equation}\label{s_value}
S = \frac{I_{[E_2,E_3]}}{I_{[E_0,E_1]}},
\end{equation}
presents a method of assessing the variability of $\sigma_n$ for a certain material, sensitive to the energy regions selected for integration. Therefore, the sharp decrease in hydrogen's cross section for neutrons with energies included in the integration regions will result in a markedly different $S$ value than that of other, low $\sigma_n(E)$-varying elements.  Fig.~\ref{fig:ii0} plots the dependence of transmission on energy, for three different materials, showing the significant difference between HDPE and metals - primarily caused by the strong energy dependence in hydrogen scattering cross section.

By choosing broadly spaced energy regions, this method mitigates the effects of material-specific $\sigma_n(E)$ resonances, such as the gradual increase of $\sigma_n(E)$ in lead from 5.5 b at 1 MeV incident neutron energy to 8 b at 4 MeV. 

\subsection{Multiple Monoenergetic Neutron Radiography}\label{sec:neutron_rad}

\begin{figure}[htbp]
\includegraphics[width=\textwidth,clip,keepaspectratio]{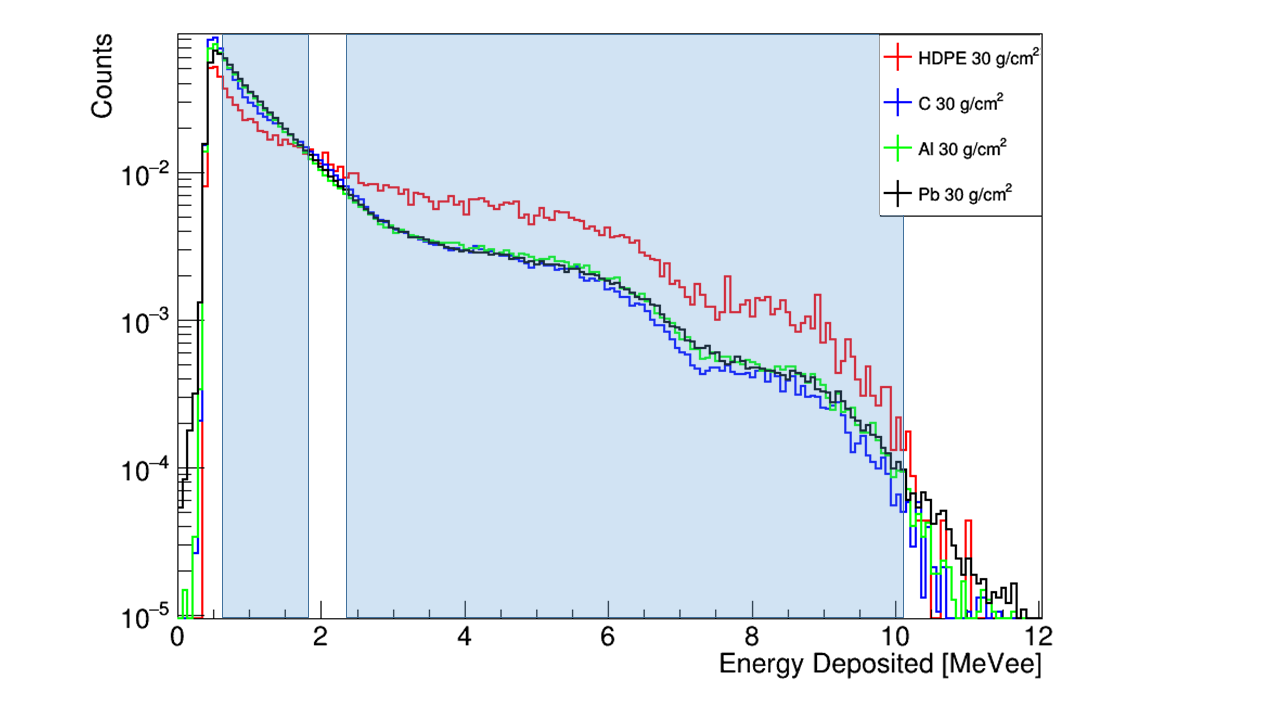}
\caption{Normalized spectra of energies deposited by transmitted neutrons from the $^{11}$B(d,n$\gamma$)$^{12}$C reaction in the experimental data. The integration regions overlaid, and is chosen based on the optimization studies described in Ref.~\cite{Rahon}. Cargoes are 30 g/cm$^2$ of HDPE, graphite, aluminum, and lead. \label{single_spec}}
\end{figure}

The 3.0 MeV deuteron used in this research generates neutron lines at the energies of 16.51, 12.2, 8.86, and 4.01 MeV, ideally suited for differential transmission analysis due to their wide energy spacing.  For a more detailed discussion of neutron production from the $^{11}B(d,n)^{12}C$ reaction see Table 1 in Ref.~\cite{ref:rahon2016spectroscopic} and Section IV in Ref.~\cite{Taddeucci}. These neutron energy groups broaden as a function of the deuteron's energy straggle within the boron target thickness. The organic scintillator detector can only detect the (partial) energy deposition of the neutrons.  Due to the dependence of the deposited energy on the neutron scattering angle, which is isotropic in the center of mass frame, the observed spectra are  continuous, monotonic distributions.  These distributions however do retain a correlation to the incident neutron energy, and as such can be used as a metric of the energy dependence of the incident neutrons. These effects result in a quasi-monoenergetic neutron beam manifesting as a series of stacked, rectangular energy distributions for each energy group, visible in Fig.~\ref{single_spec}.

The variability of the total neutron cross section for hydrogen and hydrogen-containing materials affects the energy-dependant attenuation of neutrons within the cargo volume. Examining the ratio of attenuation of high energy neutrons to those attenuated at low energy by the same sample provides information about hydrogenous content. By selecting two energy regions optimally spaced to exploit the greatest cross sectional energy dependence, the $S$ ratio described in Eqn.~\ref{s_value} becomes a representation of attenuation levels due to hydrogenous content.  Previous research has determined that the $S$ value is a statistically significant indicator of the presence of hydrogenous material~\cite{Rahon}; the work presented here demonstrates that it also has the ability to quantify and image the areal density of scanned hydrogenous material in a heterogeneous mix.

\subsection{Methods of spectral analysis and optimization}\label{sec:methods}

Without prior knowledge of the content and areal density of the cargo, a single set of integration bounds which universally generates the strongest signal of hydrogenous content must be identified. To determine these bounds, $S$ values were compared in several sets of homogeneous, non-hydrogenous transmission spectra with those of hydrogen-containing materials' transmission spectra. 
Defining $S_H$ as the attenuation measurement for the hydrogenous spectra, this number quantifies the amount of variation from the spectra of any other material, $X$, using 
\begin{equation}
R = \frac{\Bigl\lvert 1-\frac{S_X}{S_H}\Bigl\rvert}{\sigma_{\frac{S_X}{S_H}}}.
\end{equation}
This $R$ value yields units of standard deviation by dividing the absolute difference of the $S$ value ratios from unity by the propagated error, $\sigma_{\frac{S_X}{S_H}}$. For any set of integration windows, $\lbrace E_0,E_1,E_2,E_3 \rbrace$ for identically-binned energy histograms, $R$ provides a figure of merit for the analyzing power of the set. For example, $R = 2$ represents a 95\% discrimination capability between hydrogenous and non-hydrogenous materials for a certain set of integration windows. The bounds identified in this manner yield the lowest fractional error of all possible sets. 

Using an iterative $R$ value search to generate the highest discrimination of hydrogenous and non-hydrogenous cargo, an optimal set of bounds for cargo areal densities in increments of 10 \si{\gram\per\cm^2} was determined~\cite{ref:rahon2016spectroscopic}.  These bounds were applied to several experimental runs at their respective areal density and provided only a 3\% average improvement on $R$ value over the trials in which a universal, areal density-agnostic set of integration bounds were used\footnote{For the experimental setup used in this research and histogrammed spectra with 0.06 MeV bin sizes, the universal integration bounds were $\lbrace E_0, E_1 \rbrace$ = $\lbrace 10, 31 \rbrace$ and $\lbrace E_2, E_3 \rbrace$ = $\lbrace 42, 149 \rbrace$.}.

The neutron counting statistics used in this particular experimental setup limit hydrogenous material content to areal densities of $\leq$30 \si{\gram\per\cm^2}; greater hydrogenous content creates levels of neutron attenuation and energy downscatter too high to effectively employ the $R$ value spectral analysis. 


\section{Experiment}
As in previous publications regarding this work, this research was conducted at MIT's Bates Research and Engineering Center in Middleton, MA.

\begin{figure}[ht!]
    \centering
    \includegraphics[width=\textwidth,clip,keepaspectratio]{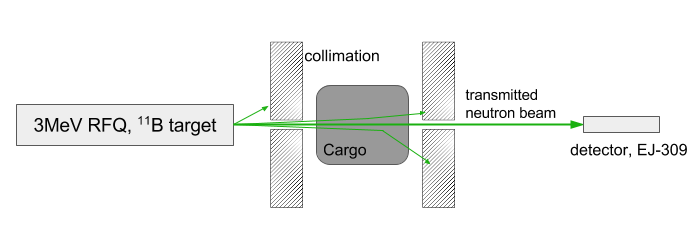}
    \caption{Top-down schematic view of the experimental radiographic layout (not to scale).}
    \label{fig:layout}
\end{figure}

An AccSys Technology Model DL-3 radio-frequency quadrupole (RFQ) linear accelerator was used to bombard a 1 mm thick natural boron target with 3.0 MeV deuterons. The $^{11}B(d,n\gamma)^{12}C$ reaction generates monoenergetic gammas of 4.4 and 15.1 MeV energies. The neutron energies were listed in Section~\ref{sec:neutron_rad}, with additional information in References~\cite{ref:rahon2016spectroscopic,Taddeucci}.
The accelerator was operated at a current of 2 to 15 $\mu$A, a beam repetition rate of 300 Hz, and a pulse length of 100 $\mu$s. Deuterons impinge on the boron target and the resulting high energy particles were collimated into a fan beam.
Various cargo phantoms with an areal density of 30 \si{\g\per\cm^2} were placed 4 m from the source target and transmitted neutrons were collimated via a second set of concrete blocks to the detector, 7 m from the cargo.  The general layout of the experiment can be seen in Fig.~\ref{fig:layout}.

A 5.08 x 5.08 cm cylindrical detector containing EJ-309 liquid organic scintillator was used for fast neutron detection. The analog waveforms were digitized via a CAEN V1720 12 bit 250 MHz digitizer. The PMT voltage was provided via a SY1527 CAEN high voltage mainframe unit. The ADAQ framework, an in-house developed software package, provides control of the digital acquisition systems and comprehensive online and offline data analysis capabilities through its integration of the ROOT toolkit~\cite{HARTWIG}.  The waveforms from the detector were analyzed with the pulse shape discrimination technique, to separate the neutron hits from the photon hits.  For a more detailed overview see Ref.~\cite{ref:rahon2016spectroscopic}, Section 3.2.

Fig.~\ref{radiograph_setup} shows a cargo schematic of a typical mixed-cargo experimental run. Four phantoms comprised of varying amounts of hydrogenous and non-hydrogenous materials were moved on a step motor-driven belt through the collimated beam line.  The hydrogenous phantoms consisted of blocks of HDPE doped with $\mathcal{O}(\%)$ fractions of boron, whose effect is ignored due to its small concentrations. Transmission neutrons and gammas were detected by the EJ-309 scintillator and neutron events isolated using the pulse shape discrimination (PSD) method. 


\section{Results}

\begin{figure}[htbp]
\includegraphics[width=\textwidth,clip,keepaspectratio]{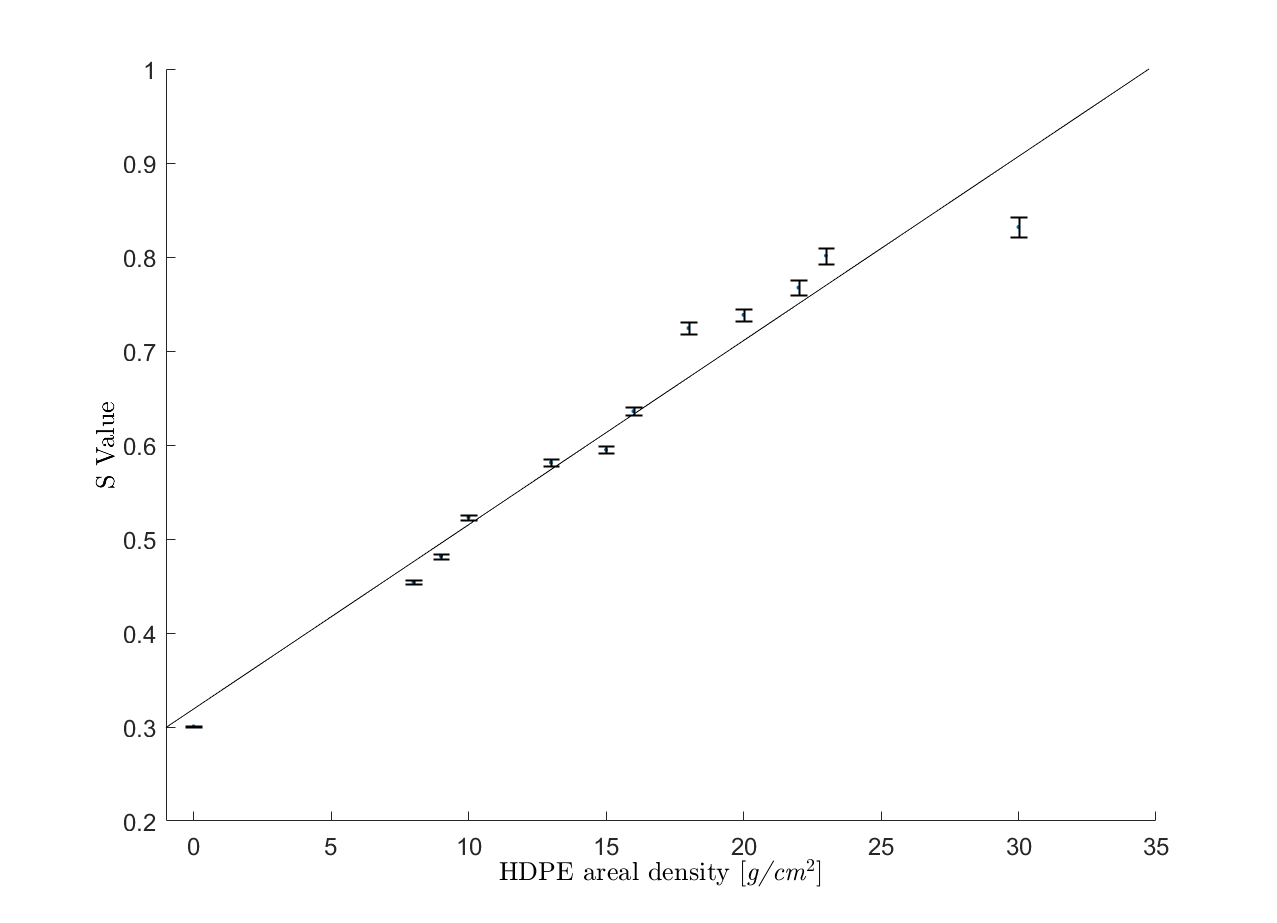}
\caption{Plot illustrating the relationship between $S$ value and areal density of HDPE in cargo. Data collected from cargoes with varying thicknesses of HDPE and lead, each cargo with an effective areal density of 30 \si{\g\per\cm^2}. Linear fit (least squares): $y=0.019x+0.33$. \label{s_val}}
\end{figure}

The $S$ value, defined in Eqn.~\ref{s_value}, is generated by applying the universal integration bounds from Sec.~\ref{sec:methods} to a given neutron spectrum. Experimental runs to determine the $S$ values of mixed hydrogenous and non-hydrogenous cargoes showed a strong dependence on HDPE content, with little effect from the additional non-hydrogenous cargo. Combined heterogeneous cargo phantoms (HDPE-aluminum, HDPE-lead, HDPE-carbon) with effective areal densities in combined thicknesses ranging from 1 \si{\gram\per\cm^2} to 50 \si{\gram\per\cm^2} were tested. The resulting plots of $S$ value to HDPE content for each material combination indicate good correlation between HDPE content and $S$ value, which is largely unaffected by the magnitude of higher-Z contributions to combined areal density. Fig.~\ref{s_val} displays the $S$ value's dependence on the actual HDPE areal density for the experimental scenarios where the combined areal density was  50 \si{\gram\per\cm^2}. 

The $S$ value's strong correlation with absolute HDPE content in a sample illustrates the predictive power of $S$ in inferring the hydrogenous content of the cargo. A linear least squares fit was applied to the data in Fig.~\ref{s_val}. The $S$ value  (which can be related to thickness, $x$, from Eqn.~\ref{eq:attn_short}) to HDPE areal density dependence is represented by an exponential relationship which can be described as $S=\frac{I_{h,0}}{I_{l,0}}\exp{[x(\Sigma_l-\Sigma_h)]}$, where the subscripts $l$ and $h$ refer to the low and high energy ranges, respectively, and the subscript $0$ refers to the initial flux.  In practice however, low energy downscatter at high HDPE ratios softens the exponential rise to be better fit by a linear relationship. 

\begin{figure}[htbp]
\includegraphics[width=\textwidth,clip,keepaspectratio]{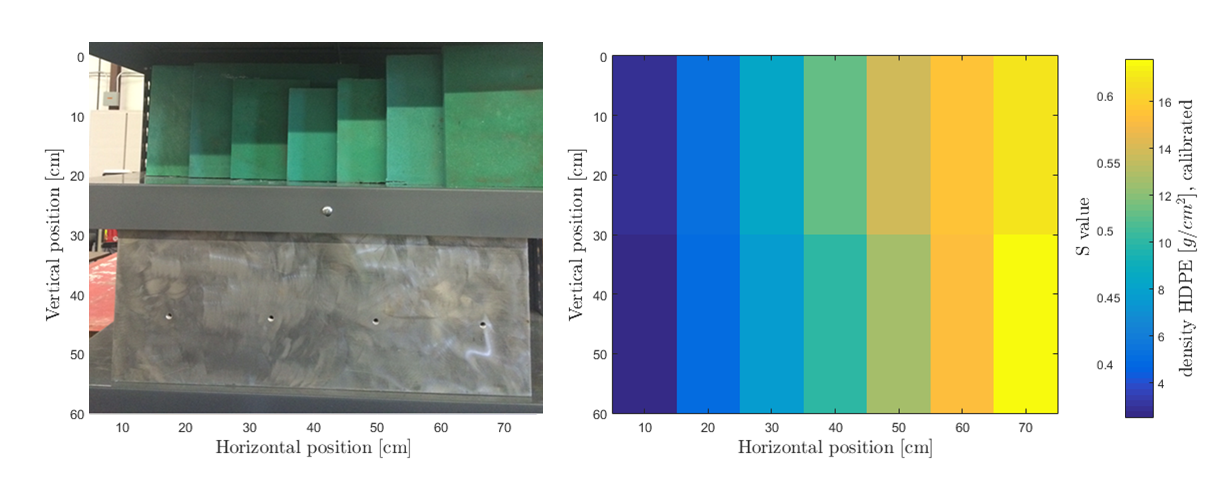}
\caption{Experimental set-up and radiograph of shielded and un-shielded HDPE
step wedges.  {\bf Left}:  a photo of a step-wedge setup of different thicknesses of HDPE without (top) and with (bottom) a 2 cm aluminum plate as shielding.  {\bf Right}: radiograph illustrating a reconstructed areal density map of the unshielded and shielded HDPE step wedges.  The data was binned into 10~cm bins, and the measurement lasted 20 minutes. \label{step_wedge}}
\end{figure}

Despite a variety of the material types and thicknesses combined with HDPE samples the $S$ value has a good linear dependence on the actual HDPE content, as shown by the fit in Fig.~\ref{s_val}. 
This indicates the strength of the correlation between HDPE quantity and $S$ value. The linear equation from the fit, $y=0.019x+0.33$ was used as a predictive model to infer HDPE areal density through observed $S$ values of cargoes of unknown composition and is the basis for the radiographic reconstruction in Fig.~\ref{step_wedge}.  Here it should be clarified that the simple linear model is purely empirical and is used here only to prove the concept.  Future research should pursue the development of more precise models and reconstruction algorithms.

The transmitted neutron count statistics became limited at areal densities greater than 30 \si{\gram\per\cm^2} of HDPE, at which point higher energy neutron attenuation and lower energy neutron down-scatter degrades the $S$ values analyzing power. Thus this work is only focused on scenarios corresponding to HDPE content of $\leq$ 30 \si{\gram\per\cm^2}.

\begin{figure}[htbp]
\includegraphics[width=\textwidth,clip,keepaspectratio]{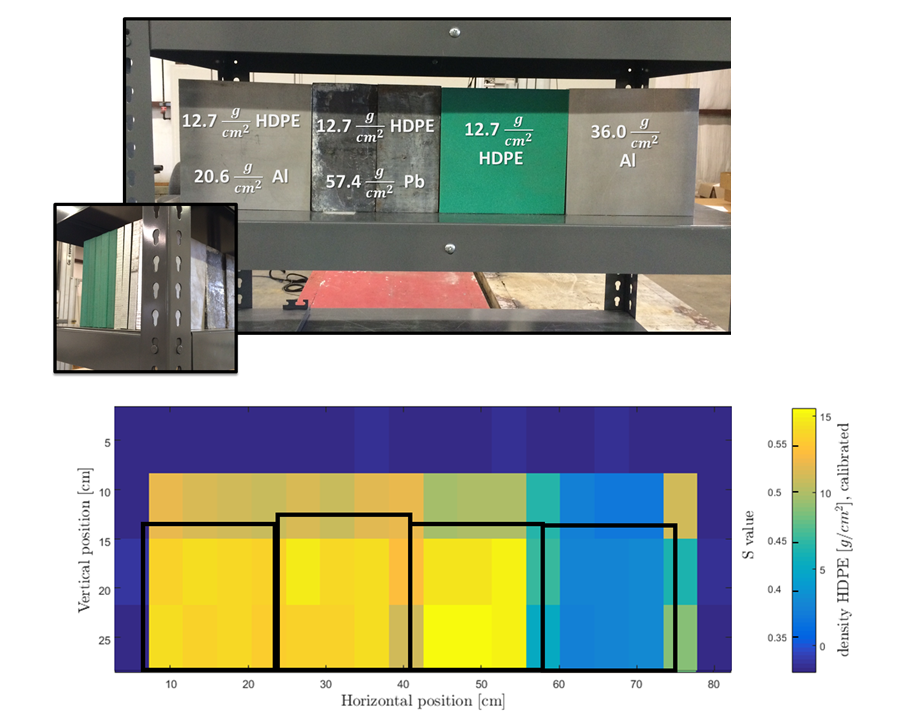}
\caption{
{\bf Top:} Cargo schematic with the areal densities overlaid.  The left two objects are a mix of metal and HDPE, the right two objects are solely HDPE and  aluminum, respectively. From left to right, total phantom areal densities are: A:33.3, B:70.1, C:12.7, and D:36.0 $\si{g/cm^2}$.  For objects A, B, and C the HDPE content was 12.7 g/cm$^2$. Inset shows sidelong view of the mock-up, revealing 12.7 $\si{g/cm^2}$ of HDPE behind the aluminum.
{\bf Bottom:} Transmission radiograph of the cargo mock-up. 
Scan performed at a speed of 0.03 $\si{cm/s}$ with a beam current of 12 $\mu$A.  Each pixel contains
a relative error of 0.3 to 1.2 percent, based on total areal density of each pixel. Phantom outlines are overlaid on the radiograph for
placement.  The vertical binning is approximately 7~cm, while the horizontal binning is 4.6~cm.
\label{radiograph_setup}}
\end{figure}

Fig.~\ref{step_wedge} shows a step-wedge cargo mock-up on the left and an S-value reconstructed radiograph on the right.  The areal density at the center of each HDPE step was inferred from the observed $S$ and the linear model described earlier. The top shelf holds a bare HDPE step wedge and the bottom shelf holds an identical step wedge, combined with 2 cm (5.4 \si{\gram\per\cm^2}) of aluminum. The accompanying radiograph and $S$ value bar scale indicate an ability to determine HDPE content to $\pm 1$ cm, with the aluminum shielding resulting in a slight underestimation of HDPE quantity. 
Fig.~\ref{radiograph_setup} shows a mixed-material cargo setup and its radiograph which is reconstructed from $S$ value. This image demonstrates the effectiveness of the radiographic technique in inferring the areal density of hydrogenous materials, even when combined with varying densities of non-hydrogenous materials.  These reconstructions show that it may be possible to use the $S$ value method of neutron radiography to provide a quantitative determination of hydrogenous content up to 30 \si{\gram\per\cm^2} of hydrogenous material for heterogeneous cargoes. Table~\ref{delta_table} shows the effectiveness of the $S$ value reconstruction technique by comparing the phantom's HDPE areal density with its calculated density.  

\begin{table}[]
    \centering
    \begin{tabular}{l l l r}
    \makecell[l]{Sample} &  \makecell[l]{Actual areal \\ density $a_0$[g/cm$^2$]} & \makecell[l]{Reconstructed average \\ areal density $a_{rec}$[g/cm$^2$]}  & \makecell[l]{Difference:\\ $(a_{rec}-a_0)/a_0$} \\
    \hline 
    A   &   12.7    &   12.3    &   -0.03 \\
    B   &   12.7    &   12.6    &   -0.01 \\
    C   &   12.7    &   13.2    &   0.043  \\
    D   &   0   &   2   &   -   \\
    \hline
    \end{tabular}
    \caption{Table comparing the areal density of the HDPE in each phantom in Fig.~\ref{radiograph_setup} and its reconstructed areal density.}
    \label{delta_table}
\end{table}


\section{Conclusions and future work}

Screening of commercial cargoes for the presence of nuclear and conventional smuggling remains a significant challenge. This work describes a novel technique which leverages the spectroscopy of fast, multiple monoenergetic neutrons to infer the hydrogenous content of the cargo which they have traversed.  The results have shown that not only is the technique sensitive to the presence of hydrogenous materials, but can also be used to accurately determine the areal density of the material\textemdash  even in heterogeneous configurations where the hydrogenous samples are masked by materials of higher atomic number, including lead.  This information can be useful both in screening cargoes for the presence of conventional contraband, as well as in the context of SNM detection.  The ability to determine the density of hydrogenous materials in the cargo can contribute towards improved specificity in SNM detection via gamma radiography techniques, as it allows to further constrain the possible values of cargo $Z$ from a non-unique observed $Z_{\mathrm{eff}}$.

The multiple monoenergetic neutron radiography technique would be most effective when combined with other radiographic techniques, including ones which include neutron detectors, such as the one described in Ref.~\cite{sowerby}.
This methodology can be extended to any other radiographic system where multiple monoenergetic neutrons are emitted and detected.  The configuration currently being developed at MIT using a 12 MeV proton cyclotron is an example of such a platforms.  Proton-induced reactions on such elements as carbon and oxygen can produce monoenergetic photons in the range of 4.4 - 10 MeV.  Simultaneously, the $^{13}C(p,n)^{13}N$ reaction on $^{13}C$, which has 1\% natural abundance, can produce monoenergetic neutrons. For a 12 MeV incident proton and a thick carbon target, the neutron energies will be at approximately 9, 6, 5 and 2 MeV. These neutrons can then be detected and their spectral analysis used to determine the hydrogenous content of the cargo.

Future research could exploit other nuclear reactions and other sources of neutrons.  While this work only focused on one type of hydrogenous cargo (HDPE) as part of a proof of concept demonstration, future experiments should include measurements on a variety of other hydrogenous materials - such as liquids, celluose (wood), and a variety of plastics.  Furthermore, more complex and mixed cargo scenarios need to be studied, and more advanced reconstruction algorithms, using broader sets of training data, can be developed.

\section{Acknowledgements}
This work is supported in part by the U.S. Department of Homeland Security Domestic Nuclear Detection Office under a competitively awarded collaborative research DE-AC52-07NA27344. It has also been supported in part by the United States Army and the United States Military Academy. The authors would like to thank their colleagues at MIT, in particular Richard Lanza and Zachary Hartwig for assistance and support.  The authors are also grateful to their collaborators at Georgia Institute of Technology and University of Michigan.  In particular the authors would like to thank Igor Jovanovich and Jason Nattress for useful discussions and comments. 

\clearpage

\section*{References}

\bibliography{neutron_transmission}

\end{document}